\documentclass[prb,showpacs,twocolumn,floatfix]{revtex4}
\usepackage{graphicx}
\usepackage{epsfig}
\usepackage{bm}
\begin{document}
\def\be{\begin{equation}}
\def\ee{\end{equation}}
\def\ba{\begin{eqnarray}}
\def\ea{\end{eqnarray}}


\title{
Interaction matrix element fluctuations in ballistic quantum dots:
dynamical effects}

\author{L. Kaplan$^1$ and Y. Alhassid$^2$}

\affiliation{\\
$^{1}$Department of Physics, Tulane University, New Orleans,
Louisiana 70118, USA \\
$^{2}$Center for Theoretical Physics, Sloane Physics Laboratory, Yale University, New Haven, Connecticut 06520, USA}

\begin{abstract}
We study matrix element fluctuations of the two-body screened Coulomb interaction and of the one-body surface charge potential in ballistic quantum dots, comparing behavior in actual chaotic billiards with analytic results previously obtained in a normalized random wave model. We find that the matrix element variances in actual chaotic billiards typically exceed by a factor of 3 or 4 the predictions of the random wave model, for dot sizes commonly
used in experiments. We discuss dynamical effects that are responsible for this enhancement. These dynamical
effects have an even more striking effect on the covariance, which changes sign when compared with random wave predictions. In billiards that do not display hard chaos, an even larger enhancement of
matrix element fluctuations is possible. These enhanced fluctuations have implications for peak spacing
statistics and spectral scrambling for quantum dots in the Coulomb blockade regime.
\end{abstract}

\maketitle

\section{Introduction}

The statistical fluctuations of single-particle energies and wave functions of dots whose single-particle dynamics are chaotic may be well approximated by random matrix theory (RMT)~\cite{guhr98}. The mesoscopic fluctuations of the conductance through open dots that are strongly coupled to leads are then successfully described by RMT~\cite{alhassid00}.  In the opposite limit of an almost-isolated dot, the charge is quantized and electron-electron interactions modify the mesoscopic fluctuations of the conductance.

The randomness of the single-particle wave functions induces randomness into the interaction matrix elements when the latter are expressed in the basis of the former. These matrix elements can be decomposed into an average and a fluctuating part. The average part of the interaction, when combined with the one-body kinetic energy and a confining potential, leads to the so-called universal Hamiltonian~\cite{kurland00,aleiner02}. This universal Hamiltonian includes a charging energy term and an exchange interaction term that is proportional to the square of the total spin of the dot (an additional Cooper-channel term is repulsive in a quantum dot and can be ignored). The fluctuating part of the interaction is suppressed by the Thouless conductance $g_T$, and in the limit $g_T \to \infty$, the dot is completely described by the universal Hamiltonian.

The charging energy term leads to charge quantization in a weakly coupled dot, and the conductance peak height distributions in such a dot were derived in Ref.~\onlinecite{jalabert92} using the RMT statistics of the single-particle wave functions. Qualitative features of these peak height distributions as well as the parametric peak height correlation and the weak localization effect as a function of magnetic field~\cite{alhassid96,alhassid98} were confirmed in experiments~\cite{folk96,chang96,folk01}. Remaining discrepancies between theory and experiments regarding the temperature dependence of the width of the peak spacing distribution~\cite{patel98a} and the peak height distributions~\cite{patel98b} at low temperatures were explained by the inclusion of the exchange interaction term of the universal Hamiltonian~\cite{alhassid03,usaj03}.

However, not all observed features of the peak spacing distribution can be explained by the exchange interaction alone. At low temperatures, the spacing is given by the second-order difference of the ground-state energy versus particle number.  When only charging energy is present, the peak spacing distribution is expected to be bimodal because of spin effects. The exchange interaction (with realistic values of the exchange coupling constant in quantum dots) reduces  this bimodality but cannot explain its {\em absence} in the experiments~\cite{sivan96,simmel97,patel98a,luscher01}. It is then necessary to consider the effect of the fluctuating part of the interaction beyond the universal Hamiltonian. 

In the Hartree-Fock-Koopmans approach, the peak spacing can be expressed in terms of certain interaction matrix elements, and sufficiently large fluctuations of such matrix elements~\cite{alhassid02} might explain the absence of bimodality in the peak spacing distribution. It is therefore of interest to make accurate estimates of interaction matrix element fluctuations in chaotic dots. These fluctuations are determined by single-particle wave function correlations. In a diffusive dot, such correlations are well understood and lead to an $O(\Delta/g_T)$ standard deviation in the interaction matrix elements~\cite{blanter97,mirlin00}, where $\Delta$ is the mean single-particle level spacing. Peak spacing fluctuations are also affected by a one-body surface charge potential induced by the accumulation of charge on the surface of the finite
dot~\cite{blanter97}. Matrix element fluctuations of the two-body interaction and one-body surface charge potential are also important for determining the statistical scrambling of the Hartree-Fock energy levels and wave functions as electrons are added to the dot~\cite{scrambling,alhassid07}.

Wave function correlations and interaction matrix elements fluctuations in a ballistic dot are less understood. In Ref.~\onlinecite{kapalh} we used a normalized random wave model to obtain analytic expressions for interaction matrix element
variances and covariances in the regime of large Thouless conductance $g_T$ for a ballistic two-dimensional dot. In such a dot, $g_T \sim kL$, where $k$ is the Fermi wave number and $L$ is the linear size of the dot (defined more precisely as the square root of the dot's area). Since $kL \sim \sqrt{N}$ where $N$ is the number of electrons in the dot, the $kL \gg 1$ limit in which the random wave model is expected to hold is also the limit of many electrons in the dot. In the present work, we systematically investigate matrix element fluctuations in real chaotic billiards, for $30 \le kL \le 70$, corresponding roughly to the parameter range relevant for experiments ($\sim 150 - 800$ electrons in the dot). We show that fluctuations can be significantly enhanced due to dynamical
effects, e.g., the variance may be enhanced by a factor of 3 or 4. Such enhancement can help in explaining the peak spacing distribution measured in the chaotic dots of Ref.~\cite{patel98a}.

On the other hand, the typical fluctuations of matrix elements in chaotic dots cannot explain the even broader peak spacing distributions in the experiment of Ref.~\onlinecite{luscher01}. The small dots used in the latter experiment are probably non-chaotic (top gates were used), and this has motivated us to study fluctuations beyond the chaotic regime.  We show that a large (i.e., order of magnitude) enhancement of the fluctuations is possible in non-chaotic billiards.

The outline of this paper is as follows. In Sec.~\ref{truechaos}, we introduce the modified quarter-stadium billiard as a convenient model for investigating matrix element fluctuations in chaotic systems. In Section~\ref{true2body}  we consider matrix elements of the two-body screened Coulomb interaction, and find strong enhancement of the fluctuations in comparison with random wave predictions.  Semiclassical corrections due to bounces from the dot's boundaries lead to an increase in the fluctuations, but do not correctly predict the scaling with $kL$ in the experimentally relevant range.  Insight into the underlying mechanism of fluctuation enhancement is obtained by studying a quantum map model, which is described in the Appendix. An important conclusion is that the expansion in $1/kL$, while asymptotically correct, can be problematic in quantifying matrix element fluctuation in the regime relevant to experiments.

In Sec.~\ref{true1body} we extend our investigation to one-body matrix elements associated with the surface charge potential,
and find similar fluctuation enhancements. Going beyond the variance, we examine the full matrix element distributions in Sec.~\ref{secstaddistr}, and observe deviations from a Gaussian shape that are even stronger than the deviations found in the random wave model~\cite{kapalh}. In Sec.~\ref{beyond} we study systems beyond the chaotic regime: billiards dominated by marginally-stable bouncing-ball modes and billiards with mixed dynamics (i.e., partly regular and partly chaotic). Finally, in Sec.~\ref{secsum} we briefly discuss some implications of the present work for the quantitative understanding of spectral scrambling and peak spacing statistics for quantum dots in the Coulomb blockade regime.

\section{Chaotic Billiards}
\label{truechaos}

\begin{figure}[ht] \begin{center} \leavevmode \parbox{0.5\textwidth}
{
\psfig{file=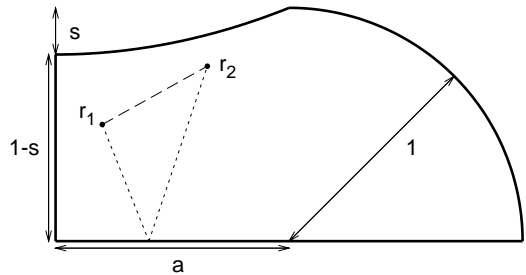,width=0.33\textwidth,angle=270}
}
\end{center}
\protect\caption{
A modified quarter-stadium geometry with parameters $a$ and $s$ is used to illustrate dynamical effects on matrix element fluctuations.  In the figure, we set the quarter-circle radius $R=1$.  The random wave contribution to the wave function intensity correlator $C(\mathbf{r_1},\mathbf{r_2})$ is schematically indicated by a dashed line, and a typical dynamical contribution by a dotted line.
}
\label{figstadpict}
\end{figure}

Here we investigate how dynamical effects modify the fluctuations of interaction matrix elements beyond our findings in the random wave model~\cite{kapalh}.
Here and in Sections \ref{true2body} -- \ref{secstaddistr}
we treat exclusively geometries displaying hard chaos.
[Systems with stable or marginally stable classical trajectories will be considered in Sec.~\ref{beyond}.]
To this end, we will
use a chaotic system shown in Fig.~\ref{figstadpict} -- a modified
quarter-stadium billiard geometry~\cite{stadium}, where the quarter-circle has radius $R$ and the straight edge of length $aR$ has been replaced by a parabolic bump to eliminate bouncing-ball modes.  Algebraically, the billiard shape is defined by
\begin{eqnarray}\label{mod-billiard}
0 \le  y /R \!\!&\le &\!\! 1-s \left(1-{x^2 \over a^2 R^2}\right) \; , \; 0
\le x/R \le a \nonumber \\0  \le  y/R \!\!&\le &\!\!\sqrt{1-(x/R-a)^2} \; ,
\; a \le x/R \le a+1 \,,
\end{eqnarray}
where $R$ is the radius of the quarter-circle, and $a$ and $s$ are free dimensionless parameters.

We use a quarter-stadium instead of a full stadium shape in order to remove symmetry effects.  This system has been verified numerically to be fully chaotic for the range of parameters used.  Variation of the bump size $s$ allows us to check the sensitivity of the results to details of the billiard geometry while maintaining the chaotic character of the classical dynamics. Furthermore,  by varying the parameter $a$, we can control the degree of classical chaos. The degree of chaos can be characterized for example by the Lyapunov exponent $\lambda$, defined as the rate of divergence at long times of generic infinitesimally separated trajectories, $|\mathbf{r}(t)-\mathbf{r'}(t)| \sim |\mathbf{r}-\mathbf{r'}| e^{\lambda t}$ as $|\mathbf{r}-\mathbf{r'}| \to 0$ and then $t \to \infty$. For $a=1.00$ and $0.1 \le s \le 0.2$, the exponent $\lambda$ takes values $0.69  \le \lambda T_B\le 0.74$ (here $T_B=mL/\hbar
k$ is a typical time scale associated with one bounce in the billiard).  When $a=0.25$, $0.55\le \lambda T_B  \le 0.56$ in the same range of $s$, indicating that the system is somewhat less chaotic for the smaller value of $a$.  Other measures of the degree of chaoticity are possible and may be more relevant to the problem of matrix element fluctuations, as we will argue below.  In particular, we may consider the rate $\lambda_\ast$ of long-time decay of classical correlations, $\overline{f(q,p)f(q(t),p(t))} - \overline{f(q,p)}^2 \sim e^{-\lambda_\ast t}$ as $t \to \infty$, where $f(q,p)$ is a typical function defined over the classical phase space and the average is over an energy hypersurface~\cite{artuso}.
Numerically, we find $0.15 \le \lambda_\ast T_B \le 0.20$ for $a=1$ and $0.095 \le \lambda_\ast T_B\le 0.13 $ for
$a=0.25$, for the same range of bump sizes $s$ as above, again indicating  a less rapid approach to ergodicity in the $a=0.25$ geometry.

An important consideration in the investigation of dynamical systems, as opposed to random wave models, is the presence of boundary conditions. Boundary conditions lead to Friedel oscillations in the average wave function intensity at distances $O(1/k)$ from a billiard boundary. The effect of such oscillations has recently been considered in Refs.~\onlinecite{tomsullmobacker}. The choice of boundary conditions, e.g., Neumann or Dirichlet, will also be seen to have significant effects on matrix element fluctuations, particularly on the fluctuations of one-body matrix elements.

Numerical wave functions for several values of the billiard parameters $a$, $s$ and in various energy ranges have been calculated  using a variation of the plane wave method~\cite{plwave}.  At each wave number $k$, a basis consisting of plane waves supplemented by a set of $Y_0$ Bessel functions centered a fraction of a wavelength outside the boundary is used; the size of the basis scales linearly with $k$.  Singular value decomposition finds at each $k$ the
linear combination that minimizes the integrated squared deviation along the boundary from the selected boundary condition (Dirichlet or Neumann).  Finally, minima of this deviation as a function of $k$ indicate the correct eigenvalues of the system.  Tests of the method include stability with respect to changes in the basis size and comparison of the resulting density of states with the Weyl formula.

Statistics are collected by averaging over an energy window.  A straightforward estimate shows that such averaging is sufficient to give good results for matrix element variances, i.e., the ratio of signal to statistical noise grows with increasing $kL$.  For all numerical results that follow, we use energy windows of constant momentum width $\Delta k\,L=10$, e.g., the data point $kL=30$ uses all wave functions within the window $25\le kL \le 35$.  The Weyl formula for the density of states in two dimensions implies that the number of wave functions in such a window grows linearly with $kL$.


\section{Two-body matrix elements}
\label{true2body}

\subsection{Fluctuation of diagonal matrix elements $v_{\alpha\beta}$}
\label{truevab}

 We first study the variance of the diagonal two-body interaction matrix elements $v_{\alpha\beta} \equiv v_{\alpha\beta;\alpha\beta}$, associated with a pair of electrons in distinct orbitals $\alpha \ne \beta$ interacting via the screened Coulomb force. Since the screening length of the Coulomb interaction in large 2D quantum dots is much smaller than the
dot size, the interaction may be modeled as
a contact interaction $v(\mathbf{r}, \mathbf{r'})=  \Delta\,V \delta(\mathbf{r}-\mathbf{r'})$, where $V=L^2$ is the dot's area, and the single-particle mean level spacing $\Delta$ serves to set the
energy scale~\cite{altshuler97,ullmo}. We then have
\begin{equation}
v_{\alpha\beta}=\Delta \, V \int_V d\mathbf{r} \, |\psi_\alpha(\mathbf{r})|^2
|\psi_\beta(\mathbf{r})|^2 \,,
\label{contactint}
\end{equation}
where the single-electron wave functions $\psi$ obey the usual normalization condition $\int_V d\mathbf{r} \, |\psi(\mathbf{r})|^2 =1$. To leading order in $1/g_T \sim 1/kL$, the variance is then given by~\cite{scrambling,kapalh}:
\begin{eqnarray}
\label{vabint}
\overline{\delta v_{\alpha\beta}^2}
= \Delta^2 V^2 \int_V \int_V d\mathbf{r} \, d\mathbf{r'} \,
C^2(\mathbf{r},\mathbf{r'}) +O\left({\Delta^2 \over (kL)^3}\right) \,,
\end{eqnarray}
where
\begin{equation}
C(\mathbf{r},\mathbf{r'}) = \overline{|\psi(\mathbf{r})|^2 |\psi(\mathbf{r'})|^2} -\overline{|\psi(\mathbf{r})|^2}
\;\overline{|\psi(\mathbf{r'})|^2}
\label{cdef}
\end{equation}
 is the intensity correlator of a single-electron wave function at points $\mathbf{r}$ and $\mathbf{r'}$. Assuming
$C(\mathbf{r},\mathbf{r'})$ is described by the normalized random-wave model (i.e., the single-electron wave functions are
normalized as above with no boundary conditions), one obtains
\begin{equation}
\overline{\delta v_{\alpha\beta}^2} =\Delta^2 {3 \over \pi} \left({2 \over \beta} \right)^2 {\ln kL
+b_g\over (kL)^2} + O\left({\Delta^2 \over (kL)^3}\right) \,,
\label{bcoeff}\end{equation}
where $\beta=1$, $2$ corresponds to the presence or absence of time reversal invariance (i.e., the absence or presence of an external magnetic field), while $b_g$ is a dimensionless coefficient that weakly depends on the dot geometry~\cite{kapalh}.

\begin{figure}[ht] \begin{center} \leavevmode \parbox{0.5\textwidth}
{
\psfig{file=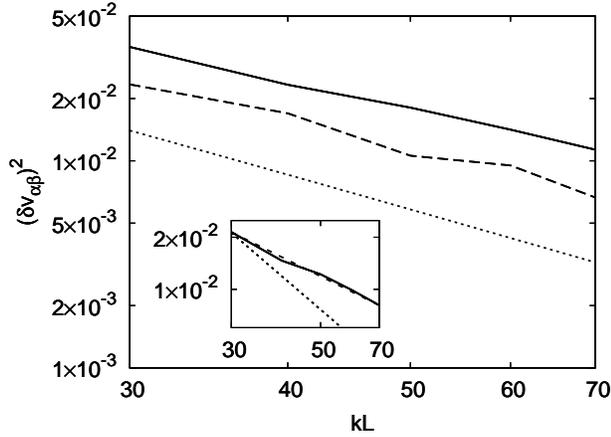,width=0.33\textwidth,angle=270}
}
\end{center}
\protect\caption{
The variance of $v_{\alpha\beta}$ versus $kL$ (on a log-linear scale) for modified quarter-stadium billiards with Neumann boundary conditions. The solid line is for $a=0.25$, while the
dashed line is for $a=1.00$. In both cases, the results are averaged over two values of the bump size: $s=0.1$ and $0.2$. Dotted line: analytic random wave prediction, Eq.~(\ref{bcoeff}).  Inset: the numerical result for $a=0.25$ with the leading logarithmic term of Eq.~(\ref{bcoeff}) subtracted (solid line) appears to fall off
as $(kL)^{-1.15}$ (dashed line).  The analytically expected subleading behavior $(kL)^{-2}$ is indicated by a dotted line for comparison.
}
\label{figstadvab}
\end{figure}

We now evaluate the variance of $v_{\alpha_\beta}$ versus $kL$ using ``exact" (numerically evaluated) real wave functions in actual chaotic billiards. Typical results are shown in
Fig.~\ref{figstadvab}, where we note the large enhancement of the billiard results over the random wave model (dotted line).  To understand this enhancement, we compare the exact numerical results for $\overline{\delta v_{\alpha \beta}^2}$ with the first term on the right hand side of Eq.~(\ref{vabint}), in which $C(\mathbf{r},\mathbf{r'})$ is taken to be the single-wave-function correlator $ C_{\rm bill}(\mathbf{r},\mathbf{r'})$  calculated numerically for the appropriate billiard system.  The discrepancy is immediately reduced to a $\sim 5-10\%$ level, which is comparable to the $O((kL)^{-3})$ higher-order correction expected and observed in the random wave model.  Thus, the large enhancement of $v_{\alpha\beta}$ fluctuations over the random wave
prediction is not due to higher-order terms in Eq.~(\ref{vabint}), but instead can be traced directly to a dynamical enhancement in the intensity correlator $C_{\rm bill}(\mathbf{r},\mathbf{r'})$ over the random-wave correlator.

 We next estimate the dynamical enhancement of the intensity correlator (as compared with a random wave model) in a semiclassical approach.  The random wave correlator $C_{\rm rw}(\mathbf{r},\mathbf{r'})$ may be interpreted semiclassically as
arising from straight-line free propagation~\cite{mirlin00} indicated by the dashed line in Fig.~\ref{figstadpict}.  As discussed by Hortikar and Srednicki~\cite{srednicki} and more recently by Urbina and Richter~\cite{urbina}, additional contributions to the correlator can be associated with trajectories that bounce off the boundary $n$ times on their way from $\mathbf{r}$ to $\mathbf{r'}$, such as the one indicated by a dotted line in Fig.~\ref{figstadpict}.  To find these contributions, we start from the
dynamical correlator for wave function amplitudes, which may be written
as~\cite{srednicki,urbina}
\begin{equation}
\overline{\psi^\ast(\mathbf{r})\psi(\mathbf{r'})} = {
\overline{G}^{\,\ast}(\mathbf{r},\mathbf{r'},E) -
\overline{G}(\mathbf{r'},\mathbf{r},E)
 \over 2\pi i\,\overline{\rho}(E)} \,,
\label{dynamp}
\end{equation}
where $\overline{G}$ is the ensemble-averaged part of the retarded Green's function
$G(\mathbf{r},\mathbf{r'},E)= \sum_\alpha {\psi^\ast_\alpha(\mathbf{r}) \psi_\alpha(\mathbf{r'})
\over E-E_\alpha +i\epsilon}$,
and $\overline{\rho}(E)$ is the smooth part of the density of states
$\rho(E)=\sum_\alpha \delta(E-E_\alpha)$.  Using Eqs.~(\ref{cdef}), (\ref{dynamp}), the dynamical intensity correlator is given by
\begin{eqnarray}
C_{\rm bill}(\mathbf{r},\mathbf{r'})= {2 \over \beta} |\overline{G}(\mathbf{r'},\mathbf{r},
E)\!-\!\overline{G}^{\,\ast}(\mathbf{r},\mathbf{r'},E)|^2 / 4\pi^2 \,\overline{\rho}^2(E)
\,.
\end{eqnarray}

 Semiclassically, the smooth density of states is given to leading order by the Weyl formula in two dimensions
\begin{equation}
\overline{\rho}(E)= {mL^2 / 2 \pi\hbar^2} \,,
\label{weyl}
\end{equation}
while the Green's function is given by the Gutzwiller
formula~\cite{gutzwiller}
\begin{equation}
\overline{G}(\mathbf{r},\mathbf{r'},E)= {1 \over i\hbar (2\pi i \hbar)^{1/2}} \sum_j |D_j|^{1/2} e^{i S_j/\hbar-i\mu_j
\pi/2}  \,. \label{gutz}
\end{equation}
The sum in (\ref{gutz}) is over classical trajectories $j$ connecting $\mathbf{r}$ to $\mathbf{r'}$ at energy $E$, $S_j$ is the action along the trajectory $j$, $\mu_j$ is the corresponding Maslov index, and $D_j$ is a classical focusing factor that scales as $m^2/p L_j$ (where $L_j$ is the length of the trajectory).  For the straight-line trajectory, $|D_j|=m^2/p|\mathbf{r}-\mathbf{r'}|$.  Inserting the semiclassical expressions (\ref{weyl}) and (\ref{gutz})
into Eq.~(\ref{dynamp}), we obtain
\begin{equation}
\overline{\psi^\ast(\mathbf{r})\psi(\mathbf{r'})} = {1 \over V} \left[J_0(k|\mathbf{r}-
\mathbf{r'}|)+ h(\mathbf{r},\mathbf{r'}) (kL)^{-1/2}\right] \,,
\label{ampdyncorr}
\end{equation}
where the Bessel function arises from the straight-line path,
and $h(\mathbf{r},\mathbf{r'})$ is a sum over all other trajectories:
\begin{eqnarray}
h(\mathbf{r},\mathbf{r'})\!&=&\!\! \sum'_j h_j(\mathbf{r},\mathbf{r'}) \nonumber \\ \!&=&\!\! \sum'_j
\left|{2 pL D_j \over \pi m^2}\right|^{1 \over 2}\!\!\cos\left (\!{S_j \over \hbar}\!-\!
{(2\mu_j+1)\pi \over 4}\!\right) \,.  \label{hsum}
\end{eqnarray}
For typical point pairs $(\mathbf{r},\mathbf{r'})$ separated by a distance of order $L$, the function $h(\mathbf{r},\mathbf{r'})$ is order unity in $kL$, and the contributions to the correlator from the straight line path and from other paths are both $O((kL)^{-1/2})$.
For pairs $(\mathbf{r} ,\mathbf{r'})$ separated by a bouncing path of length $L_j/L \le \epsilon \ll 1$, $h(\mathbf{r},\mathbf{r'}) \sim \epsilon^{-1/2}$.  However, the fraction of such pairs is $O(\epsilon^3)$ and their contribution to the variance and other moments of matrix element distributions is negligible.

The intensity correlator in the semiclasssical approximation becomes
\begin{eqnarray}
C_{\rm sc}(\mathbf{r},\mathbf{r'}) &=& {1 \over V^2}{2 \over \beta}
\left[J_0^2(k|\mathbf{r} -\mathbf{r'}|)  +h^2(\mathbf{r}
,\mathbf{r'}) (kL)^{-1}\right.  \nonumber \\ &+& 2J_0(k|\mathbf{r}- \mathbf{r'}
|) h(\mathbf{r},\mathbf{r'}) (kL)^{-1/2}  \left. \right] \,,
\label{cbill}
\end{eqnarray}
where the first (random wave) term is associated with the straight-line path, and the remaining terms constitute
semiclassical corrections.

Similarly to the random wave correlator~\cite{gornyi02,urbina07,kapalh},
$C_{\rm sc}(\mathbf{r},\mathbf{r'})$ must be corrected to take into account individual wave function normalization. In analogy with Refs.~\onlinecite{gornyi02,kapalh} we have, to leading order in $1/kL$,
\begin{eqnarray}
& & \tilde C_{\rm sc}(\mathbf{r},\mathbf{r'}) = C_{\rm sc}(\mathbf{r},\mathbf{r'}) +
{1 \over V^2} \int_V \int_V d\mathbf{r_a} d\mathbf{r_b} \,C_{\rm sc}(\mathbf{r_a},\mathbf{r_b}) \nonumber \\
& &- \,{1 \over V}
\int_V d\mathbf{r_a} \, C_{\rm sc}(\mathbf{r},\mathbf{r_a}) - {1 \over V} \int_V d\mathbf{r_a} \, C_{\rm sc}(\mathbf{r_a},\mathbf{r'}) \,.
\label{corrnormbill}
\end{eqnarray}
Substituting $\tilde C_{\rm sc}$ for $C$ in
(\ref{vabint}), we find
\begin{equation}
\overline{\delta v_{\alpha\beta}^2}  \!= \! \Delta^2 {3 \over \pi} \left({2 \over \beta} \right)^2 {(\ln kL
+b_g)+b_{\rm sc}\over (kL)^2} + O\left({\Delta^2 \over (kL)^3}\right) \,,
\label{bsc}
\end{equation}
where $b_{\rm sc}$ is a classical constant that in practice must be determined numerically by performing the integral in Eq.~(\ref{vabint}).  As noted above, the random wave and semiclassical contributions to $ C_{\rm sc}(\mathbf{r}
,\mathbf{r'})$ are of the same order except for $|\mathbf{r}-\mathbf{r'}| \ll L$; it is
these short-distance pairs that result in a logarithmic enhancement of the random-wave term.

We may easily estimate the dependence of $b_{\rm sc}$ on the degree of chaoticity of the dynamical system by invoking a diagonal approximation, in which the intensity correlator $C_{\rm sc}(\mathbf{r},\mathbf{r'})$ of Eq.~(\ref{cbill}) is averaged over
classically small regions surrounding $\mathbf{r}$ and $\mathbf{r'}$.  Noting that Eq.~(\ref{hsum}) gives $h(\mathbf{r},\mathbf{r'})$ as a sum of oscillatory terms with quasi-random phases, such averaging leads to
\begin{eqnarray}
C_{\rm sc}^{\rm dg}(\mathbf{r},\mathbf{r'})\! =\!{1 \over V^2}{2 \over \beta}
\big[J_0^2(k|\mathbf{r} \!-\!\mathbf{r'}|)
 \!+\!{1 \over kL}\!\sum'_j h_j^2(\mathbf{r},\mathbf{r'}) \big] \,,
\label{cbilldiag}
\end{eqnarray}
where $\sum'_j h_j^2(\mathbf{r},\mathbf{r'})$ corresponds to the total classical probability of traveling from a neighborhood of $\mathbf{r}$ to a neighborhood of $\mathbf{r'}$ via paths $j$ other than the straight-line path.  Naively, the average semiclassical correction to the intensity correlator appears to increase as we include longer trajectories.  However, let us organize the trajectories by number of bounces $n$ or by time $t \sim nT_B$,
where $T_B$ is a typical time for one bounce in the billiard.  Trajectories at times $t$ that are significantly longer than the classical correlation decay time $\lambda_\ast^{-1}$ contribute only a constant, independent of $\mathbf{r}$ and $\mathbf{r'}$, to $C_{\rm sc}^{\rm dg}(\mathbf{r},\mathbf{r'})$. This is because a classical cloud of trajectories centered near $\mathbf{r}$ becomes approximately equidistributed over the entire billiard when $e^{\lambda_\ast t} \gg 1$, for any initial point $\mathbf{r}$.  Such position-independent contributions to $C_{\rm sc}^{\rm dg}(\mathbf{r},\mathbf{r'})$ get subtracted off in the normalization procedure (\ref{corrnormbill}).  Thus, the typical size of $ C_{\rm sc}(\mathbf{r},\mathbf{r'})$ is determined by trajectories $j$ having no more than $n_{\rm max} \approx (\lambda_\ast T_B)^{-1}$ bounces.

Furthermore, as a function of $t$, the number of classical trajectories typically grows as $e^{\lambda t}$, while the focusing factor for each trajectory $j$ falls off as $|D_j| \sim e^{-\lambda t}$, where $\lambda$ is the Lyapunov exponent defined earlier.  Thus, all $n$-bounce trajectories combine to form a contribution to Eq.~(\ref{cbilldiag}) whose order is roughly $n$-independent for $n <n_{\rm max}$.  Summing over $n$ up to $n_{\rm max}$, where $n_{\rm max}$ is large, we find
\begin{equation}
 C_{\rm sc}^{\rm dg}(\mathbf{r},\mathbf{r'})={1 \over V^2}{2 \over \beta}
\left[J_0^2(k|\mathbf{r} -\mathbf{r'}|) + O\left ({n_{\rm max} b_1 \over
kL}\right)\right]\,, \label{diagscale}
\end{equation}
where $b_1$ characterizes the size of the semiclassical contribution from one-bounce trajectories.  Going beyond the diagonal approximation is necessary to evaluate properly the integral in Eq.~(\ref{vabint}), but the scaling is unaffected.  Comparing Eqs.~(\ref{vabint}), (\ref{bsc}), and (\ref{diagscale}), we obtain an estimate for the coefficient $b_{\rm sc}$ in Eq.~(\ref{bsc}) describing the semiclassical correction to the random wave model
\begin{equation}
b_{\rm sc} \sim n_{\rm max}^2 b_1^2 \sim \left({ b_1 \over \lambda_\ast T_B
}\right)^2\,. \label{bscscale}
\end{equation}
This estimate confirms our intuition that semiclassical corrections to the random wave approximation become increasingly important as we consider billiards with a very long ergodic time $\lambda_\ast^{-1}$.

Alternatively, the scaling (\ref{bscscale}) may be obtained by noting that when classical correlations persist on a time scale $\lambda_\ast^{-1}$ that is much longer than the one-bounce time $T_B$, then the effective dimensionless Thouless conductance is reduced to $g_T \sim (\lambda_\ast T_B) kL$. Now a typical chaotic wave function $\psi_\alpha(\mathbf{r})$ may be written as a
superposition of $O(g_T)$ non-ergodic basis states $\eta_i(\mathbf{r})$.
Since the
correlator $\overline{\eta_i^\ast(\mathbf{r})\eta_i(\mathbf{r'})}$ for each non-ergodic
basis state $\eta_i$ is of order $V^{-1}$, we easily see that $\overline{\psi_\alpha^\ast(\mathbf{r}) \psi_\alpha(\mathbf{r'})}$
takes typical values of order $V^{-1} g_T^{-1/2}$.  The wave function intensity correlator $C_{\rm sc}(\mathbf{r},\mathbf{r'})$ scales as the square of the amplitude correlator, or as $V^{-2} g_T^{-1}$ for typical pairs $(\mathbf{r},\mathbf{r'})$, yielding a lower bound
\begin{equation}
\overline{\delta v_{\alpha\beta}^2} \sim {\Delta^2 \over g_T^2}
\sim {\Delta^2 \over (\lambda_\ast T_B kL)^2}
\end{equation}
for the integral (\ref{vabint}), consistent with Eqs.~(\ref{bsc}) and (\ref{bscscale}).

 For ``generic" chaotic systems, the correlation decay time $\lambda_\ast^{-1}$ is of the same order as the one-bounce time $T_B$, and the above scaling arguments for $\lambda_\ast T_B \ll 1$ are not applicable.  Instead, only the first few bounces may contribute in practice to the semiclassical correlator, but these must be summed up numerically to obtain the semiclassical
coefficient $b_{\rm sc}$.  This coefficient may in practice be quite large even for generic chaotic systems (e.g., the modified stadium billiard) and grows as the system becomes less chaotic.


Qualitatively, the above discussion is consistent with our billiard results shown in Fig.~\ref{figstadvab}, as fluctuations are observed to be consistently larger for the less chaotic $a=0.25$ billiard, as compared with the $a=1.00$ billiard.  We note that both billiards are ``generic", in the sense that they are not fine-tuned to obtain an anomalously long time scale $\lambda_\ast^{-1}$.  We also note that varying the bump size $s$ has a very weak effect on the matrix element statistics (as long as $s$ is large enough to
destroy the bouncing-ball modes) and serves instead to provide an estimate of the statistical uncertainty in our results.

\begin{figure}[ht] \begin{center} \leavevmode \parbox{0.5\textwidth}
{
\psfig{file=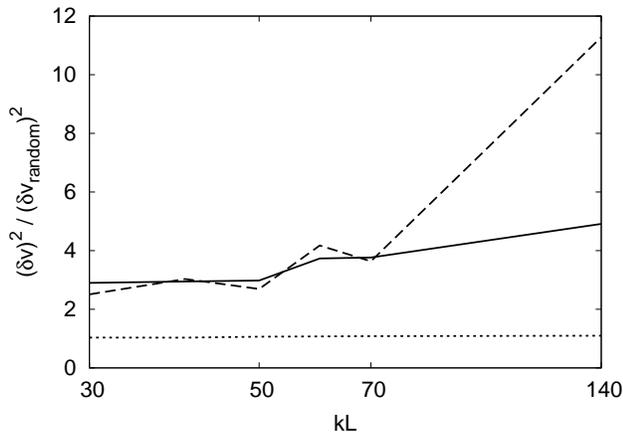,width=0.33\textwidth,angle=270}
}
\end{center}
\protect\caption{
The enhancement of the variance of $v_{\alpha\beta}$ (solid line),
$v_{\alpha\alpha}$ (dashed line) and $v_{\alpha\beta\gamma\delta}$ (dotted
line) over the corresponding random wave predictions  is shown for $a=0.25$ billiards. [For $v_{\alpha\beta}$, the random wave prediction is given by Eq.~(\ref{bcoeff}), and analogous
expressions for the other matrix elements may be found in Ref.~\onlinecite{kapalh}.]  In
each case, the data is averaged over bump sizes $s=0.1$ and $0.2$.
}
\label{figstadvaaratio}
\end{figure}

 For the modified quarter-stadium billiard, we have found that adding
one-bounce effects to the random wave correlator increases the predicted
$v_{\alpha\beta}$ variance by $\sim 30$ -- $40\%$ in the energy range of interest, a
significant change but not nearly sufficient to explain the full factor of $3$
-- $5$ enhancement observed in Fig.~\ref{figstadvaaratio} for
the $a=0.25$ billiards (solid lines).  Indeed, a close look at the data suggests that the
numerical results cannot be explained fully by semiclassical arguments, no
matter how many bounces are included in the analysis.  The semiclassical
correction to the variance in Eq.~(\ref{bsc}) is manifestly $O(1/(kL)^{2})$.
However, the inset in
Fig.~\ref{figstadvab} clearly shows that the dynamical
contribution to the variance with $kL$ is not consistent with Eq.~(\ref{bsc})
but instead appears to follow a much slower power law $\sim 1/(kL)^{-1.15}$.
This may be seen also in Fig.~\ref{figstadvaaratio} (solid line), where the enhancement over
the random wave prediction grows instead of diminishing with increasing
$kL$.

 We believe this anomalous behavior results from a combination of two
related factors: the dynamical enhancement, discussed above, of the $b_{\rm sc}$
coefficient due to a finite correlation time scale $\lambda_\ast^{-1}$ in an
actual dynamical system, and the consequent saturation of the $1/(kL)^2$ behavior at moderate
($\alt 100$) values of $kL$.  As the classical system becomes less unstable and
the correlation time $\lambda_\ast^{-1}$ increases, $b_{\rm sc}$ also increases
in accordance with Eq.~(\ref{bscscale}), leading to greatly enhanced matrix
element variance at very large values of $kL$ (\ref{bsc}).  Because the
variance is bounded above independent of $kL$, the $(kL)^{-2}$ growth in the
variance necessarily breaks down for smaller values of $kL$.  This
small-$kL$ saturation sets in at ever larger values of $kL$ as the system
becomes less unstable and $\lambda_\ast^{-1}$ becomes larger.

 Alternatively,
one may note that the natural expansion parameter for interaction matrix
element fluctuations in a dynamical system is not $(kL)^{-1}$ but rather the
inverse Thouless conductance $g_T^{-1} \sim (\lambda_\ast T_BkL)^{-1}$,
and the semiclassical contribution with prefactor $b_{\rm sc}$
in Eq.~(\ref{bsc}) is the leading $O(g_T^{-2})$ effect in such an expansion.
Terms of third and higher order in $g_T^{-1}$, although formally subleading and
not included in a semiclassical calculation, become quantitatively as large as
the leading $O(g_T^{-2})$ term when $g_T$ falls below some characteristic
value.  Furthermore, if one considers chaotic billiards with a long correlation
decay time $\lambda_\ast^{-1}$, the importance of formally subleading terms in
the $g_T^{-1}$ expansion will extend to quite large values of $kL$.

\begin{figure}[ht] \begin{center} \leavevmode \parbox{0.5\textwidth}
{
\psfig{file=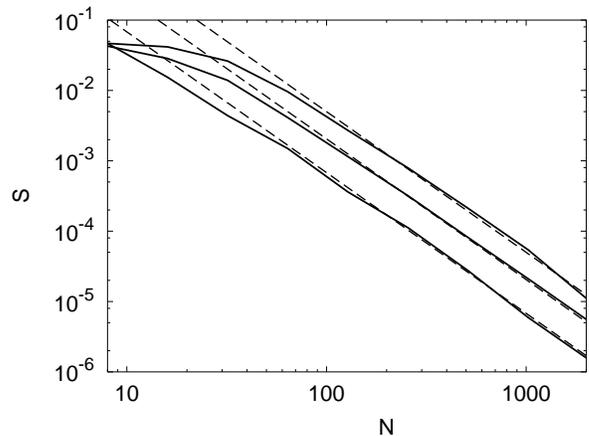,width=0.33\textwidth,angle=270}
}
\end{center}
\protect\caption{
The two-body matrix element variance $S$ for a quantum map, Eq.~(\ref{maps}) in the Appendix, as a function of the Hilbert space dimension $N$.  From top to bottom, the three solid lines represent data for dominant orbit stability exponent $\lambda_0=0.25$, $0.50$, $1.00$.  The three dashed lines indicate the asymptotic $1/N^2$ behavior for each case in the semiclassical regime of large $N$.  }
\label{figmap}
\end{figure}

 The above assertions are explicitly confirmed for a quantum map model,
described in detail in the Appendix, which has scaling behavior analogous to
that of a two-dimensional billiard, with the number of states $N=2\pi/\hbar$
playing the role of semiclassical parameter $kL=pL/\hbar$ in the
billiard~\cite{map1,map2}.  As in the billiard, a free parameter in the
definition of the map allows for control of the classical correlation decay
time $\lambda_\ast^{-1}$.  A key difference between the two-dimensional
billiard and the map model is that the map lacks a logarithmic random wave
contribution to the variance.  We see in Fig.~\ref{figmap} that the expected
$N^{-2}$ behavior of the variance is observed at sufficiently large $N$, for
all three families of quantum maps considered.  Furthermore, the prefactor multiplying
$N^{-2}$ in each case agrees with that obtained from a semiclassical
calculation, and as expected this prefactor grows with increasing classical
correlation time $\lambda_\ast^{-1}$ (corresponding to a decrease in the
chaoticity of the system).  We also see in Fig.~\ref{figmap} that even for a
``typical" chaotic system (i.e., $\lambda_\ast T_B \sim 1$), strong deviations
from the $1/N^2$ law appear already below $N \approx 80$.  Such deviations
extend to even larger $N$ for chaotic systems with slower classical correlation
decay.  This suggests that the large-$N$ or large-$kL$ expansion, though
theoretically appealing and asymptotically correct, is problematic in
describing the quantitative behavior of interaction matrix element fluctuations
for real chaotic systems in the physically interesting energy range.

The above numerical calculations were all performed in the presence of time
reversal symmetry ($\beta=1$). From Eq.~(\ref{bsc}) we see
that when time reversal symmetry is broken ($\beta=2$), both the random wave contribution to the matrix element
variance (the term proportional to $\ln kL +b_g$) and the semiclassical contribution (the term proportional to $b_{\rm sc}$) are suppressed by the same factor of $4$. Thus, the dynamical enhancement factor for a given dot geometry
is necessarily $\beta$-independent in the semiclassical limit $kL \gg 1$. However, the saturation effect, which tends to suppress the enhancement as $kL$ is reduced, will be less important when $\beta=2$, since
the variance is smaller in this case. Thus, at any finite value of $kL$, the dynamical enhancement in the variance over the random wave model will be greater when time reversal symmetry is broken, and one may expect enhancements
somewhat larger than those shown in Fig.~\ref{figstadvaaratio}. This result has been confirmed in the quantum map model.

\subsection{Fluctuation of $v_{\alpha\alpha}$ and
$v_{\alpha\beta\gamma\delta}$}

 We have similarly studied the variance $\overline{\delta v_{\alpha\alpha}^2}$ of double-diagonal interaction matrix elements and the variance $\overline{\delta
v_{\alpha\beta\gamma\delta}^2}$ of off-diagonal interaction matrix elements for actual chaotic billiards.  Once again, the random wave predictions~\cite{kapalh} must be used as the baseline for comparison.  In Fig.~\ref{figstadvaaratio}, we show the enhancement factor for these matrix element variances, together with the corresponding data for $\overline{\delta v_{\alpha\beta}^2}$ discussed previously.

 In the range $30 \le kL \le 70$ most relevant to experiment, we observe an
enhancement in $\overline{\delta v_{\alpha\alpha}^2}$ over the random wave
prediction that is similar to the enhancement in $\overline{\delta
v_{\alpha\beta}^2}$ in the same energy range.  In both cases, the enhancement
factor continues to grow, instead of approaching unity, at increasing $kL$.
This latter fact strongly suggests that even at $kL=140$, we are still far from
the asymptotic regime of large $g_T$, where matrix element fluctuations would
be adequately described by a random wave picture supplemented by semiclassical
corrections.  The enhancement at large $kL$ is particularly dramatic in the
case of $\overline{\delta v_{\alpha\alpha}^2}$ fluctuations.  On the other hand, the
variance of off-diagonal matrix elements $v_{\alpha\beta\gamma\delta}$ is
enhanced over the random wave prediction by at most 10\%, over the entire
energy range considered.  This is consistent with the reasonable expectation
that dynamical effects lead to particularly strong deviations from random wave
behavior in a modest fraction of the total set of single-particle states, such
as those associated with particularly strong scarring on unstable periodic
orbits~\cite{hellerscar}.  Such deviations lead to a significant tail in the
$v_{\alpha\alpha}$ distribution, but have a minimal effect on the distribution
of off-diagonal matrix elements, since it is unlikely for all four wave
functions $\psi_\alpha$, $\psi_\beta$ $\psi_\gamma$, and $\psi_\delta$ to be
strongly scarred or antiscarred on the same orbit.

 Indeed, inspection of wave functions $\psi_\alpha$ associated with anomalously
high double-diagonal matrix elements $v_{\alpha\alpha}$ shows that these wave
functions have disproportionately high intensity on average near the dominant
horizontal bounce periodic orbit, which follows the lower edge of the billiard
in Fig.~\ref{figstadpict}.  We note, however, that asymptotic scar theory in
the $kL \to \infty$ limit predicts $O(1/(kL))$ corrections to the intensity
correlation function in position space and only in a region of size
$O(1/(kL)^{1/2})$ surrounding a periodic orbit.  Comparing with the integral
expression (\ref{vabint}) for the variance, we see that periodic orbits
asymptotically contribute to the variance only at order $1/(kL)^{3}$, compared
to the $O(1/(kL)^{2})$ semiclassical effect associated with generic
(non-periodic) classical trajectories (\ref{bsc}).  Thus, the relative
importance of periodic orbit effects on matrix element fluctuations is a
finite-$kL$ (or finite-$\hbar$) phenomenon, which cannot explain the
quantitative scaling behavior of the variance with $kL$, and which is expected
to become irrelevant in the asymptotic $kL \to \infty$ limit.

\subsection{Matrix element covariance $\overline{\delta v_{\alpha \beta}
\delta v_{\alpha\gamma}}$}
\label{seccovaractual}

 The normalized random wave model has been shown to produce a covariance $\overline{\delta v_{\alpha \beta}
\delta v_{\alpha\gamma}}$ that is always negative, has size $\sim\Delta^2 \ln kL/(kL)^3$ for small $\omega=E_\beta - E_\omega$, and falls off as $(\omega/E_T)^{-2} \sim (\delta k L)^{-2}$ for $\omega \gg E_T$, where $E_T$ is the ballistic Thouless energy~\cite{kapalh}.  However, in a diffusive
dot, the same matrix element covariance is found to be a {\it positive} constant $\propto \Delta^2/g_T^3$ (where $g_T$ is the diffusive Thouless conductance) for energy separations $\omega$ much smaller than the diffusive Thouless energy $E_c$.  This diffusive covariance falls off for $\omega \gg E_c$ but remains positive as long as $\omega \ll \hbar /\tau$, where $\tau$ is the mean free time \cite{scrambling}.  An interesting issue is then the sign of the covariance in an actual chaotic system.

First we note the sum rule~\cite{kapalh}
 \begin{equation}\label{covar-var}
\sum_{\beta \neq \gamma} \overline{\delta v_{\alpha\beta} \delta
v_{\alpha\gamma} } = -\sum_\beta  \overline{(\delta v_{\alpha\beta})^2} \;.
\end{equation}
This sum rule is quite general and holds for either a ballistic or a diffusive dot as long as a completeness relation is satisfied within an energy window in which the states $\beta$ and $\gamma$ reside.  The average covariance must therefore be negative when averaged over all states $\beta$ and $\gamma$ within such an energy
window.  The size of the energy window in each case must be at least of size $\hbar$ multiplied by the inverse time scale of first recurrences.  In a ballistic system this implies an energy window of size at least $E_0=\hbar/T_B$, where $T_B$ is the one-bounce time.  In a diffusive system, the completeness relation requires energy scales larger than $E_0=\hbar/\tau$, where $\tau$ is the mean free time, and thus the positive sign of the diffusive
covariance at energy separations $\omega \ll \hbar/\tau$ does not contradict the sum rule (\ref{covar-var}).

In actual chaotic billiards, it is in principle possible to find positive covariance at energy scales $\omega \ll E_0$, as long as the covariance is sufficiently negative for $\omega \sim E_0$ to produce a negative average covariance over the full energy window that is consistent with the sum rule (\ref{covar-var}).  Such positive covariance can result from scars since $\psi_\beta$ and $\psi_\gamma$ will typically be scarred or antiscarred along the same orbits when $\omega = E_\beta - E_\gamma$ is small.  The scar contribution to the covariance for small $\omega$ is $O(1/(kL)^3)$ (i.e., of the same order as the scar contribution to the variance)
and is formally subleading compared with the negative $O(\ln kL/(kL)^3)$ random wave contribution.  However, within the range of $kL$ values relevant to experiments, the scar contribution can dominate and lead to a positive covariance for nearby single-particle wave functions.

\begin{figure}[ht] \begin{center} \leavevmode \parbox{0.5\textwidth}
{
\psfig{file=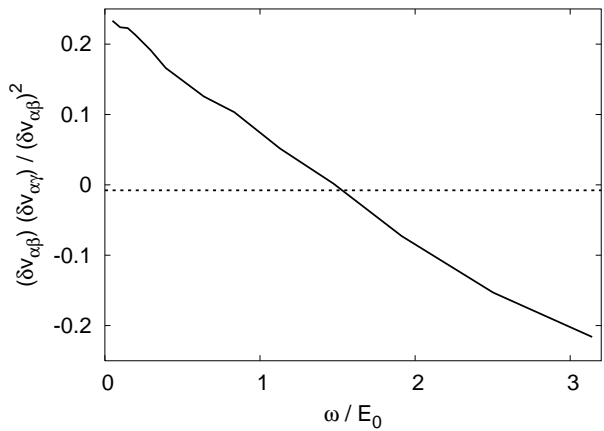,width=0.33\textwidth,angle=270}
}
\end{center}
\protect\caption{
 The covariance $\overline{\delta v_{\alpha \beta} \delta v_{\alpha\gamma}}$ is
computed as a function of energy separation $\omega=E_\beta-E_\gamma$ for an
ensemble of ballistic discrete-time maps, described in the Appendix,
Eqs.~(\ref{vqrnd}) and (\ref{kprnd}).  Here $E_0=\hbar/T_B$, where $T_B$ is the one-bounce time.  The system size $N$ is $128$, and $A=0$.  The dotted line indicates the negative average covariance implied by the sum rule (\ref{covar-var}).}
\label{figcovarmap}
\end{figure}

Unfortunately, it is not practical to calculate the matrix element covariance in a real billiard, since the number of wave functions that can be averaged over is not sufficient to obtain a signal larger than the statistical noise. We instead obtain good statistics for the covariance in a ballistic discrete map model,
introduced previously in the discussion of the variance, and described in detail in the Appendix.  In such discrete maps, the matrix element variance or covariance contains no logarithmic terms.  For generic chaotic ballistic systems
(i.e., Lyapunov time of the same order as the one-step time), we
find that the covariance is $O(N^{-3}) \sim O((kL)^{-3})$ and positive for $\omega \ll E_0 =\hbar/T_B$, but becomes negative at $\omega \sim E_0$, in contrast with the random wave prediction of an always negative covariance.  A typical example for $N=128$ is shown in Fig.~\ref{figcovarmap}.  Here discreteness of time implies energy periodicity with period $2\pi E_0=2\pi \hbar/T_B$, and thus a maximum energy separation $\omega=\pi E_0$.  In Fig.~\ref{figcovarmap}, the dotted line indicates the negative average covariance over the entire energy window of size $2\pi E_0$, as required by the sum rule (\ref{covar-var}).

\begin{figure}[ht] \begin{center} \leavevmode \parbox{0.5\textwidth}
{
\psfig{file=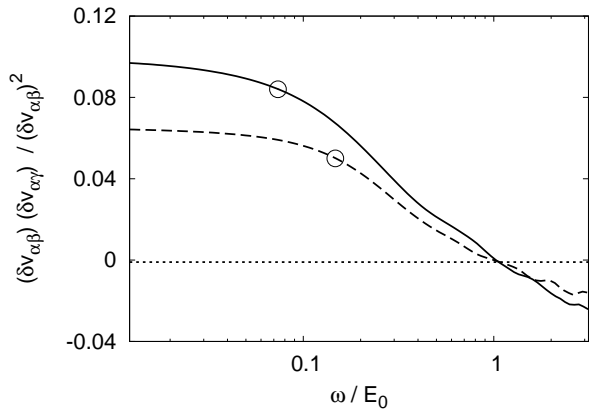,width=0.33\textwidth,angle=270}
}
\end{center}
\protect\caption{
The covariance $\overline{\delta v_{\alpha \beta} \delta v_{\alpha\gamma}}$ is computed as a function of energy separation $\omega=E_\beta-E_\gamma$ for an
ensemble of diffusive discrete-time maps on a 32x32 lattice~\cite{ossipov}. The solid curve corresponds to Thouless conductance $g_T=12$ ($E_c/E_0=0.074$) and the dashed curve corresponds to $g_T=24$ ($E_c/E_0=0.147$). Here $E_0=\hbar/\tau$, where $\tau$ is the mean
free time.
The value $\omega=E_c$, below which the covariance is expected to approach a constant positive value,
is indicated by a circle in each case.
The dotted line indicates the negative average covariance implied by the sum rule (\ref{covar-var}).
}
\label{figcovardiff}
\end{figure}

 It is interesting to compare with the covariance in an ensemble of
two-dimensional diffusive discrete maps~\cite{ossipov}.  Typical data is shown
in Fig.~\ref{figcovardiff} for an ensemble of diffusive maps on a 32x32
lattice, with Thouless conductance $g_T=12$ (solid curve) and $g_T=24$ (dashed
curve).  The theory predicts a variance scaling as $1/g_T^2$ and a covariance
scaling as $1/g_T^3$, so $\overline{\delta v_{\alpha \beta} \delta
v_{\alpha\gamma}}/\overline{\delta v_{\alpha \beta}^2}$ should scale as $1/g_T$
in the $g_T \to \infty$ limit.  Just as in the ballistic case, the covariance
is positive for small separations $\omega$ and becomes negative when $\omega
\sim E_0$.  The average covariance over a maximal energy window of size $2\pi
E_0$ is again negative, as predicted by the sum rule (\ref{covar-var}) and
indicated by a dotted line.

\section{One-body matrix elements}
\label{true1body}

When an electron is added to the finite dot, charge accumulates on the surface
and its effect can be described by a one-body potential energy ${\cal V}(\mathbf{r})$.
The diagonal matrix elements of ${\cal V}(\mathbf{r})$ are given by
$
v_\alpha \equiv {\cal V}_{\alpha \alpha} =\int_V d
\mathbf{r} \; |\psi_\alpha(\mathbf{r})|^2 \, {\cal V}(\mathbf{r})
$,
and the variance of these one-body matrix elements may be computed as
\begin{equation}
\label{v1expr}
\overline{\delta v_\alpha^2} = \int_V \int_V d\mathbf{r} \, d\mathbf{r'} \; {\cal
V}(\mathbf{r}) C(\mathbf{r},\mathbf{r'}) {\cal V}(\mathbf{r'})  \,.
\end{equation}
Dynamical enhancement of one-body matrix element fluctuations may be studied similarly to the analysis of two-body matrix element fluctuations presented in Sec.~\ref{true2body}.
The leading semiclassical contribution to the variance is obtained
by substituting the normalized semiclassical intensity correlator
$C_{\rm sc}^{\rm dg}$ [see Eq.~(\ref{diagscale})] for $C(\mathbf{r},\mathbf{r'})$ in Eq.~(\ref{v1expr}).  We immediately obtain
\begin{equation} \overline{\delta v_\alpha^2} ={ c_g +c_{\rm sc} \over \beta}
{\Delta^2 \over kL} +O\left({\Delta^2 \over (kL)^2}\right) \,,
\label{sconebody}
\end{equation}
where $c_g$ is a geometry-dependent dimensionless coefficient arising already
in the random wave model~\cite{kapalh}, while $c_{\rm sc} \sim (\lambda_\ast T_B)^{-1}$ is associated with
the classical dynamics.  We note that the asymptotic power-law behavior of the
variance is unchanged from the random wave model, and the variance is enhanced
only by a $kL$-independent constant.

\begin{figure}[ht] \begin{center} \leavevmode \parbox{0.5\textwidth} {
\psfig{file=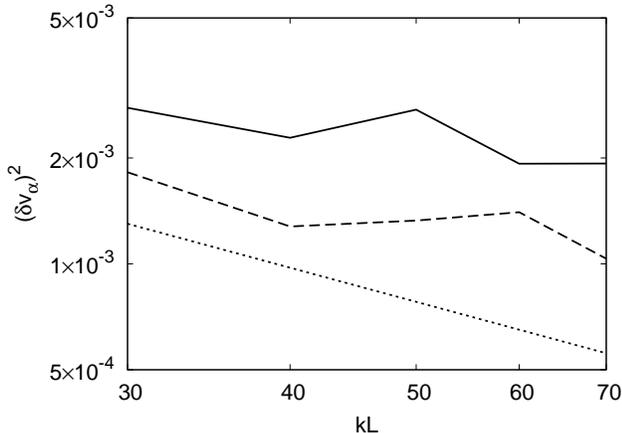,width=0.33\textwidth,angle=270}
}
\end{center}
\protect\caption{
The variance of the one-body diagonal matrix element $v_{\alpha}$ for modified quarter-stadium billiards ($a=0.25$; averaged over $s=0.1$ and $s=0.2$) is plotted as a function of semiclassical parameter $kL$.  Solid line: Neumann boundary conditions.  Dashed line: Dirichlet boundary conditions on curved boundaries, and Neumann boundary conditions elsewhere. Dotted line: Analytic prediction for the random wave model (given by Eq.~(\ref{sconebody}), including only the $c_g$ term).
}
\label{figstadva}
\end{figure}

\begin{figure}[ht] \begin{center} \leavevmode \parbox{0.5\textwidth}
{
\psfig{file=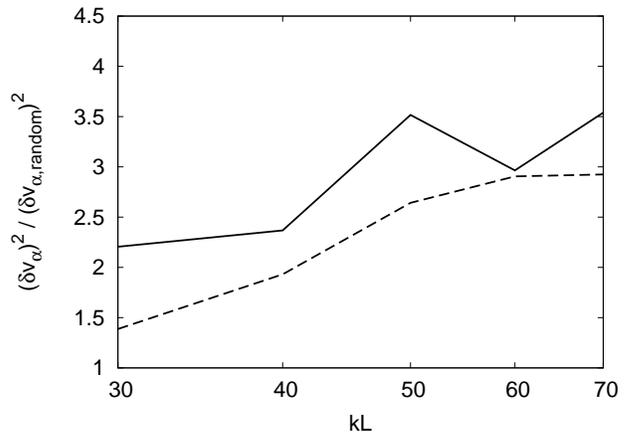,width=0.33\textwidth,angle=270}
}
\end{center}
\protect\caption{
Enhancement factor of the $v_\alpha$ variance over the random wave prediction
is plotted for modified quarter-stadium billiards
with Neumann boundary conditions, averaged over $s=0.1$ and $0.2$.  Solid line: $a=0.25$;
dashed line: $a=1.00$.
}
\label{figstadvaratio}
\end{figure}

Numerical data for $\overline{\delta v_\alpha^2}$ in modified quarter-stadium
billiards is presented in Fig.~\ref{figstadva}, and compared with random wave
results.  The ratio of the actual variance to the random wave prediction is
shown in Fig.~\ref{figstadvaratio}.  Clearly this ratio is not constant but
rather grows with $kL$ (as was also the case with the $v_{\alpha\beta}$
variance), indicating once again that at $kL \approx 70$ we have not yet
reached the asymptotic large-$kL$ regime where semiclassical expressions become
applicable.  The same can be observed by comparing data for Neumann and
Dirichlet boundary conditions in Fig.~\ref{figstadva}.  Since Dirichlet wave
functions decay to zero at distances less than $1/k$ from a boundary, where the
surface potential is especially strong, we expect larger matrix element
fluctuations for the Neumann boundary condition data, qualitatively consistent with the results in
the figure.  However, the fraction of points $\mathbf{r}$ so close to the boundary
is $O(1/kL)$, while the surface potential ${\cal V}(\mathbf{r})$ is only enhanced
by $O((kL)^{1/2})$ there, so the boundary condition effect is formally
subleading.  Nevertheless, we clearly see from the figure that in the energy
range of experimental interest, the boundary condition effect is of size
comparable both to the dynamical enhancement and to the baseline random wave
prediction for the variance.

\section{Matrix element distributions}
\label{secstaddistr}

\begin{figure}[ht] \begin{center} \leavevmode \parbox{0.5\textwidth}
{
\psfig{file=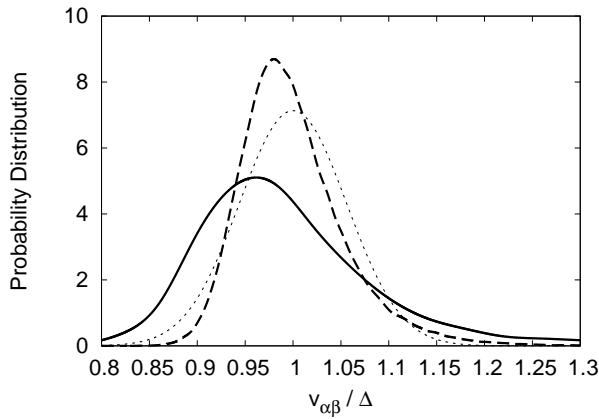,width=0.33\textwidth,angle=270}
}
\end{center}
\protect\caption{
The distribution of diagonal interaction matrix elements $v_{\alpha\beta}$ is
shown for real random waves in a disk~\cite{kapalh} (dashed curve) and for actual eigenstates
in a modified quarter-stadium billiard geometry with Neumann boundary conditions
(solid curve) at $kL=70$.  A Gaussian distribution with the same mean
and variance as the random wave distribution is shown as a dotted curve for
comparison.
}
\label{figdistr}
\end{figure}

 Just as was done previously for the random wave model~\cite{kapalh}, we can go beyond the
variance to investigate higher moments of the matrix element distribution for
actual chaotic systems.  A typical distribution for diagonal two-body
matrix elements $v_{\alpha\beta}$ in a modified quarter-stadium billiard with
$a=0.25$ and $s=0.1$ is shown in Fig.~\ref{figdistr}.  Since the approach to
Gaussian behavior is already very slow in the case of random waves, it is not
surprising to find even stronger deviations from a Gaussian shape for matrix
elements in real chaotic systems at the same energies.  Thus, for modified
quarter-stadium billiards with $a=1$, the skewness $\gamma_1$ of the
$v_{\alpha\beta}$ distribution {\it grows} from $1.95$ at $kL=70$ to $2.72$ at
$kL=140$, while the skewness for the same geometry in the random wave model
drops slightly from $1.21$ to $1.09$.  Similarly, the excess kurtosis
$\gamma_2$ increases from $8.3$ at $kL=70$ to $20.9$ at $kL=140$, while
dropping from $3.7$ to $3.3$ in the random wave model.  Similar behavior is
obtained for other matrix elements.  Clearly, the distribution tails are very
long, and the assumption of Gaussian matrix element distributions is even less
justified for real chaotic systems than it was in the random wave model.

\section{Beyond the Chaotic Regime}
\label{beyond}

 In this Section we consider fluctuations of matrix elements in systems that
are not fully chaotic.  Here no universal behavior is expected but we shall see
that in such systems the variance can be enhanced much more than in fully
chaotic systems~\cite{ullmo}.  We use the modified quarter-stadium billiard
[see Eq.~(\ref{mod-billiard})] with $s=0$ or $a<0$.  The choice $s=0$
corresponds to the original Bunimovich stadium, whose quantum fluctuation
properties are dominated by the marginally-stable bouncing-ball modes, while
$a<0$ corresponds to a lemon billiard, which has a classically mixed,
or soft chaotic, phase space.

\subsection{Two-body matrix elements}

\subsubsection{Fluctuation of diagonal matrix elements $v_{\alpha\beta}$}
\label{secvabmixed}

 In contrast with the $\ln kL/ (kL)^2$ falloff in the $v_{\alpha\beta}$
variance predicted for fully chaotic dynamics by Eq.~(\ref{bsc}), in the case
of regular or mixed dynamics we expect $kL$-independent matrix element
fluctuations of order unity.  To see this explicitly, suppose that the
classical phase space consists of one regular and one chaotic region, with each
wave function uniformly distributed over one of the two regions.  Projecting
these regions onto position space, let $f(\mathbf{r})$ be the fraction of the
energy hypersurface at $\mathbf{r}$ that is part of the regular region, i.e., the
fraction of momentum directions at $\mathbf{r}$ that correspond to stable
trajectories.  Then the average regular wave function has intensity
$\overline{|\psi_{\rm reg}(\mathbf{r})|^2} =V^{-1} f(\mathbf{r})/\overline{f}$ at
position $\mathbf{r}$, while the average chaotic wave function has intensity
$\overline{|\psi_{\rm ch}(\mathbf{r})|^2} =V^{-1} (1-f(\mathbf{r}))/(1-\overline{f})$.
Here
$
\overline{f}={1 \over V} \int_V d\mathbf{r_a} \; f(\mathbf{r_a})
$
is the total fraction of regular points in classical phase space, or
equivalently the fraction of regular quantum eigenstates in the large $kL$
limit.  Then, starting with the expression (\ref{contactint}) for the two-body
matrix element we find that on average
\begin{equation}
v_{\alpha\beta} = \Delta V \int_V d\mathbf{r} \; {1 \over V^2}
{f^2(\mathbf{r}) \over {\overline f}^2}=
\Delta { \overline{f^2} \over {\overline f}^2}
\end{equation}
whenever $\alpha$ and $\beta$ are both regular states, to be compared with the
overall average $\overline{v_{\alpha\beta}}=\Delta$ for {\it all} states
$\alpha$, $\beta$.  Clearly, $v_{\alpha\beta}$ is enhanced by a factor of order
unity, since the two regular states tend to be concentrated in the same region
of phase space.  Similarly, by replacing $f$ with $1-f$, we obtain enhanced
$v_{\alpha\beta}= \Delta (\overline{f^2}-2\overline{f}+1)/(1-\overline{f})^2$
when both $\alpha$ and $\beta$ are chaotic, and finally, below average
interaction matrix elements $v_{\alpha\beta}= \Delta
(\overline{f^2}-\overline{f})/(\overline{f}^2-\overline{f})$
are typically obtained when one single-particle state is regular and the other
chaotic.  Combining these results, we obtain the lower bound
\begin{equation}
\overline{\delta v_{\alpha\beta}^2} \ge \Delta^2 \left(
{\overline{f^2}-\overline{f}^2 \over \overline{f}-\overline{f}^2  } \right)^2
\,,
\label{mixedbound}
\end{equation}
where the quantity in parentheses is a classical system property independent of
$kL$.  Unless the local regular fraction $f(\mathbf{r})$ is a position-independent
constant, this quantity is nonzero, and the standard deviation is necessarily
of the order of $\Delta$, i.e. of the same order as the average
$v_{\alpha\beta}$.  We note that Eq.~(\ref{mixedbound}) is a lower bound only,
as it assumes that each regular or chaotic state is uniformly spread over its
corresponding phase space region.  Any intensity fluctuations within the set of
regular states or within the set of chaotic states will only add to the total
matrix element variance.

 The $kL$-independence of the variance can also be inferred from the following simple argument: regular-like quantum behavior is obtained when the ergodic
time $\lambda_\ast^{-1}$ becomes of the same order as the Heisenberg time $\pi
kL T_B$ needed to resolve the spectrum.  Then the Thouless conductance $g_T
\sim kL \lambda_\ast T_B$ is of order unity and Eqs.~(\ref{bsc}) and
(\ref{bscscale}) imply $ \overline{\delta v_{\alpha\beta}^2} \sim \Delta^2$.

 The constant factor in Eq.~(\ref{mixedbound}) depends not only on the regular
fraction $\overline{f}$ in phase space, but equally importantly on the relative
size $\sim \overline{f}^2/\overline{f^2}$ of the position-space region in which
the regular states live (i.e., the participation ratio of the regular states).
For example, in the extreme case where all regular
states live in in area $V_{\rm reg}$ and all chaotic states live in the
complementary area $V-V_{\rm reg}$, we have $\overline{f}=\overline{f^2}=V_{\rm
reg}/V$, and $\overline{\delta v_{\alpha\beta}^2}=\Delta^2$, independent
of the size of $V_{\rm reg}$.

 Eq.~(\ref{mixedbound}) predicts very large enhancement, scaling as
$(kL)^2/\ln kL$, of the matrix element variance in mixed dynamical systems,
over the random wave prediction.  Large matrix element fluctuations
in the presence of soft chaos have previously been observed in
Ref.~\onlinecite{ullmo}.

\begin{figure}[ht] \begin{center} \leavevmode \parbox{0.5\textwidth}
{
\psfig{file=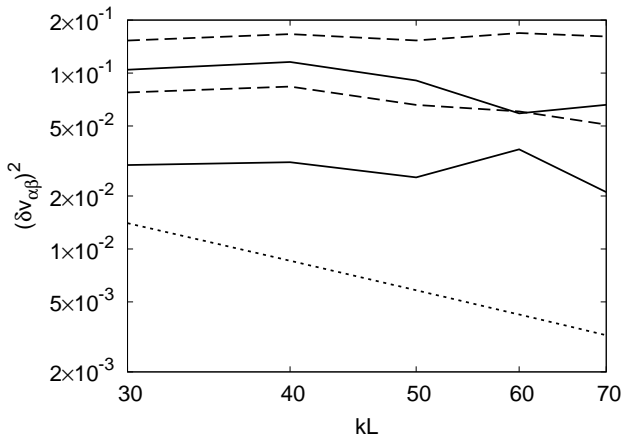,width=0.33\textwidth,angle=270}
}
\end{center}
\protect\caption{
The variance of $v_{\alpha\beta}$ for $a=0.25$, $1.00$ quarter-stadium billiards
(upper and lower solid lines); $a=-0.25$, $-0.50$ quarter-lemon billiards
(upper and lower dashed lines); random waves (dotted line).  Neumann boundary
conditions are used for all four billiards.  }
\label{figvabmix}
\end{figure}

\begin{figure}[ht] \begin{center} \leavevmode \parbox{0.5\textwidth}
{
\psfig{file=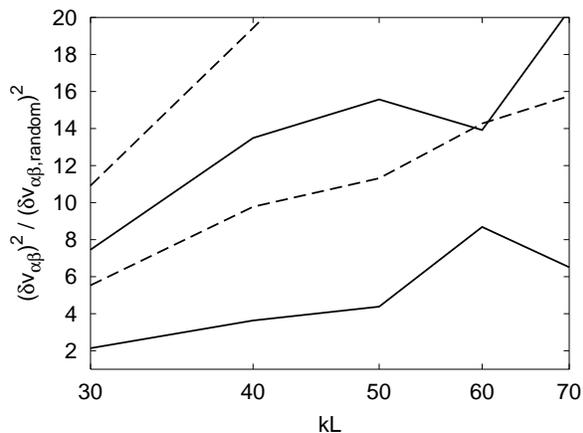,width=0.33\textwidth,angle=270}
}
\end{center}
\protect\caption{
Enhancement of the $v_{\alpha\beta}$ variance as compared with the random wave prediction for $a=0.25$, $1.00$ quarter-stadium billiards (solid lines);
$a=-0.25$, $-0.50$ quarter-lemon billiards (dashed lines).  See
Fig.~\ref{figvabmix}.  }
\label{figvabmixratio}
\end{figure}

 The diagonal matrix element variance $\overline{\delta v_{\alpha\beta}^2}$ is
computed as a function of $kL$ for two typical mixed phase-space quarter-lemon
billiards and shown by dashed lines in Fig.~\ref{figvabmix}.  As expected, no
falloff with $kL$ is observed.  In Fig.~\ref{figvabmixratio}, we see that
enhancement of an order of magnitude or more over random wave behavior can
easily be obtained for physically interesting values of $kL$.  The most
dramatic enhancement is observed for the $a=-0.25$ quarter-lemon billiard,
which is closer to integrability.

Behavior intermediate between hard chaos and mixed chaotic/regular phase space
is obtained in the presence of families of marginally stable classical
trajectories, such as the ``bouncing ball" orbits of the stadium billiard.
In the quarter stadium billiard ($s=0$ in Fig.~\ref{figstadpict}),
exceptional states associated with such orbits are concentrated in the
rectangular region of the billiard and constitute a fraction $\sim
1/(kL)^{1/2}$ of the total set of states~\cite{baecker}.  When $\alpha$ and
$\beta$ are both bouncing ball states, $\delta v_{\alpha\beta}=v_{\alpha\beta}-
\overline{v_{\alpha\beta}} \sim \Delta$, just as would be the case for regular
states concentrated in a finite fraction of the available coordinate space.
These special matrix elements dominate the variance, leading to
\begin{equation}
\overline{\delta v_{\alpha\beta}^2} \sim {\Delta^2 \over kL} \,,
\label{bouncebound}
\end{equation}
and implying an enhancement factor $\sim kL /\ln kL$ over the random wave
prediction.  Numerical data for quarter-stadium billiards is shown by solid
lines in Figs.~\ref{figvabmix} and \ref{figvabmixratio}.  The stronger
fluctuations are observed in the less chaotic $a=0.25$ stadium.

\subsubsection{Fluctuation of $v_{\alpha\alpha}$ and
$v_{\alpha\beta\gamma\delta}$}

 A calculation analogous to the one resulting in Eq.~(\ref{mixedbound}) shows
that $\overline{\delta v_{\alpha\alpha}^2}$ must also be $O(\Delta^2)$ and
$kL$-independent for a billiard with mixed phase space.  In addition, the
average $\overline{v_{\alpha\alpha}}$ is enhanced by an $O(1)$ factor from its
random wave value of $3\Delta$ ($\beta=1$) or $2\Delta$ ($\beta=2$).  In the
stadium billiard, the absence of a stable phase space region ensures that
bouncing ball states, with $\delta v_{\alpha\alpha} \sim \Delta$ and frequency
$\sim 1/(kL)^{1/2}$ should dominate the double-diagonal matrix element
variance:
\begin{equation}
\overline{\delta v_{\alpha\alpha}^2} \sim {\Delta^2 \over (kL)^{1/2}} \,.
\end{equation}
The billiard results (not shown) are qualitatively consistent with the above predictions, although statistical noise prevents us from extracting a meaningful power law behavior.

 In contrast, fluctuations in the off-diagonal matrix elements
$v_{\alpha\beta\gamma\delta}$ are relatively little affected by bouncing ball
orbits or even regular phase space regions.  This is due to the fact that these
elements are zero on average, not $O(\Delta)$, and thus an increase by an
$O(1)$ factor of some matrix elements does not necessarily lead to a large
variance.  We may consider an extreme scenario where each eigenstate is located
in one of two disjoint regions of area $V/2$.  Clearly
$v_{\alpha\beta\gamma\delta}$ is non-vanishing only when all four states are
located in the same half of the billiard.  In such a case, the typical
$v_{\alpha\beta\gamma\delta}^2$ is enhanced by a factor of $8$ compared
with the random wave prediction, ignoring logarithms.
Because $1/8$ of all matrix elements
$v_{\alpha\beta\gamma\delta}$ are nonzero, the variance $\overline{\delta
v_{\alpha\beta\gamma\delta}^2}$ is nearly unchanged from the ergodic case.  The
above argument generalizes trivially to an arbitrary number of wave function
classes.  Numerical data in quarter-stadium and quarter-lemon billiards (not
shown) confirm that $\overline{\delta v_{\alpha\beta\gamma\delta}^2}$ is nearly independent of the classical dynamics in the billiard.  Higher moments of the $\delta v_{\alpha\beta\gamma\delta}$ distribution are greatly enhanced in systems with mixed phase space, and the distribution becomes strongly non-Gaussian.

\subsection{One-body matrix elements}

\begin{figure}[ht] \begin{center} \leavevmode \parbox{0.5\textwidth}
{
\psfig{file=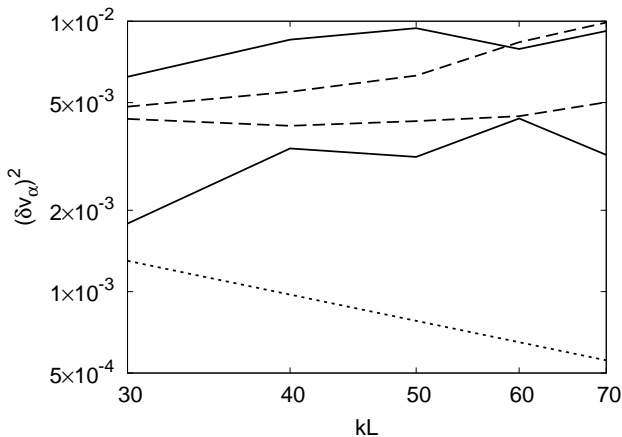,width=0.33\textwidth,angle=270}
}
\end{center}
\protect\caption{
The variance of $v_{\alpha}$ for $a=0.25$, $1.00$ quarter-stadium billiards
(solid lines); $a=-0.25$, $-0.50$ lemon billiards (dashed lines); random waves
(dotted line).  Neumann boundary conditions are used for all four billiards.
}
\label{figvamix}
\end{figure}

 In a billiard with mixed classical phase space, we expect the one-body matrix
element $v_\alpha$ of the surface charge potential ${\cal V}$ to average
$\int_V d\mathbf{r} \; {\cal V}(\mathbf{r})f(\mathbf{r})/\int_V d\mathbf{r} \; f(\mathbf{r}
)=\overline{{\cal V} f}/\overline{f}$ for regular states, where $f(\mathbf{r})$ is
the function defined in Section~\ref{secvabmixed}, and similarly to average
$(\overline{\cal V}-\overline{{\cal V}f})/(1-\overline{f})$ for chaotic states.
We then obtain a lower bound for the variance analogous to
Eq.~(\ref{mixedbound}),
\begin{equation}
\overline{\delta v_{\alpha}^2} \ge  { \left(
\overline{{\cal V}f}-\overline{\cal V}\;\overline{f} \right)^2 \over \overline{f}-
\overline{f}^2}
\,,
\label{mixedboundva}
\end{equation}
which is $O(\Delta^2)$ and independent of $kL$.  Thus, Eq.~(\ref{mixedboundva})
implies an enhancement by a factor $\sim kL$ over the variance for fully
chaotic billiards given by Eq.~(\ref{sconebody}).  The absence of a falloff in
the variance with increasing $kL$ is consistent with our results for quarter-lemon billiards (dashed lines) in Fig.~\ref{figvamix}.

 In the quarter-stadium billiard, bouncing-ball states with $\delta v_{\alpha} \sim \Delta$ will once again dominate the variance
\begin{equation}
\label{vabounce}
\overline{\delta v_{\alpha}^2} \sim {\Delta^2 \over (kL)^{1/2}}
\,,
\end{equation}
which is a factor $\sim (kL)^{1/2}$ enhancement over random wave behavior. The decay predicted by Eq.~(\ref{vabounce}) is not observed in the numerical data
in the experimentally relevant range $30 \le kL \le 70$ (solid lines in Fig.~\ref{figvamix}), suggesting once again that the energies are not high enough for the asymptotic large-$kL$ scaling laws to be applicable.
We do find that enhancement by a factor of $5$ to
$15$ of the one-body matrix element variance is quite possible in the energy
range of interest, when the billiard under consideration exhibits either soft
chaos or marginally stable orbits in the classical dynamics.


\section{Summary and conclusion}
\label{secsum}

We have studied fluctuations of two-body and one-body matrix elements in chaotic billiards as a function of a semiclassical parameter $kL$, and compared them with the normalized random wave model predictions. Understanding the quantitative behavior of these fluctuations is important for the proper analysis of peak spacing statistics in the Coulomb blockade regime of weakly coupled chaotic quantum dots.

Dynamical effects, associated with non-random short-time behavior in actual chaotic systems, are formally subleading for two-body matrix
elements, and of the same order as the random wave prediction for one-body matrix elements. In practice, however, we find that these effects can easily lead to enhancement by a factor of $3$ or $4$ of the variance in both one-body and two-body matrix elements for experimentally relevant values of $kL$ and in reasonable hard chaotic geometries.
Somewhat larger enhancement factors are expected when time reversal symmetry is broken by a magnetic field. 
The size of these dynamical corrections scales in each case as a power of $\lambda_\ast^{-1}$, a time scale associated with
approach to ergodicity in the associated classical dynamics.  Random wave behavior is recovered in the limit $\lambda_\ast^{-1} \to 0$. In typical geometries, dynamical effects on matrix element fluctuations cannot be properly computed in a semiclassical approximation, as higher-order terms are quantitatively of the
same size as the semiclassical expression in the $kL$ range of experimental interest. We have used a quantum map model to investigate the approach to semiclassical scaling at very large values of $kL$ as well as the saturation of matrix element fluctuations at moderate to small values of $kL$.

 In the case of the interaction matrix element covariance for energy levels that are separated by less
than the ballistic Thouless energy, dynamical effects are not only
often larger than random wave effects, but are also of opposite sign, leading to an overall covariance that is positive. This is in contrast with the random wave model where the covariance is always negative.  Nevertheless, the sum rule (\ref{covar-var}) is preserved due to large negative covariances for more widely separated states.  We have discussed an analogy with similar behavior in diffusive systems.

Systems with a mixed chaotic-regular phase space or with families of marginally stable classical orbits show even stronger enhancement of matrix element fluctuations as compared with the random wave model.  We discussed the expected asymptotic scaling with $kL$ of the matrix element fluctuations in these cases, and found it to be very different from the scaling found in chaotic systems.

Our results strongly indicate that wave function statistics in actual chaotic single-particle systems, including dynamical effects, are needed to make a proper quantitative comparison between theory (e.g., Hartree-Fock) and experiment.  A better understanding of single-particle wave function correlations is then essential for the calculation of observables in an interacting many-electron system such as the peak spacing distribution in the Coulomb blockade regime of a quantum dot.  Furthermore, these correlations need to be understood beyond the naive leading order semiclassical approximation, to allow comparison with experiments, which are generally performed at moderate values of the semiclassical parameter $kL$.

\section*{Acknowledgments} We acknowledge useful discussions with Y.~Gefen, Ph.~Jacquod, and C.\ H.~Lewenkopf.  This work was supported in part by the U.S. Department of Energy Grants No.\ DE-FG03-00ER41132 and DE-FG-0291-ER-40608 and by the National
Science Foundation under Grant No.\ PHY-0545390. We are grateful for the hospitality of the Institute for Nuclear Theory
at the University of Washington, where this work was completed.

\appendix \section{Quantum Map Model}

 To understand better the anomalously slow decay of $\overline{\delta v_{\alpha\beta}^2}$ and other matrix element fluctuations in realistic chaotic
systems, we may consider a toy model (perturbed cat map~\cite{pertcat}) that
displays very similar behavior and for which it is easy to collect good
statistics at very large values of $kL$.  Define a classical map on the torus
$(q,p) \in [-\pi,\pi) \times [-\pi,\pi)$ by
\begin{eqnarray}
q_{t+1} &=& q_t +K'(p_t)\;\; {\rm mod}\;\; 2\pi \nonumber \\
p_{t+1}&=&p_t-V'(q_{t+1}) \;\; {\rm mod} \;\; 2\pi \,.
\end{eqnarray}
The above map may be obtained by stroboscopically viewing the
periodically-kicked Hamiltonian system
\begin{eqnarray}
H(q,p,t)=K(p)+\sum_{n=-\infty}^\infty \delta(t-n) V(q) \,.
\end{eqnarray}
We choose the kick potential to be a perturbation of an inverted harmonic
oscillator
\begin{eqnarray}
V(q)&=&-{q^2 \over 2} - A\cos q -B(4\cos q- \cos{2q}) \nonumber \\ &+& C
(2\sin q -\sin{2q}) \,,\end{eqnarray}
while the kinetic term governing free evolution between kicks is
\begin{equation}
K(p)={p^2 \over 2} + A\cos p +B (4\cos p-\cos{2p}) \,.
\end{equation}
$K(p)$ is even in $p$ to preserve a time-reversal invariance (symmetry class
$\beta =1$).  $V(q)$ and $K(p)$ have been chosen so that the map has a period-1
orbit at $q=p=0$, with stability exponent
\begin{equation}
\lambda_0 = \cosh^{-1}\left[1+ {1 \over 2}(1-A)^2\right] \approx 1-A \,,
\end{equation}
where the approximate form holds for $\lambda_0 \ll 1$.  Thus, $A$ may be
varied to change the stability of the shortest orbit, whereas the perturbations $B$
and $C$, which have no effect on the linearized behavior around $q=p=0$, allow
for ensemble averaging while keeping the monodromy matrix of the central orbit
fixed.

 This map may be quantized using standard techniques~\cite{map1}; the position
basis is discrete with spacing $\hbar$ due to periodicity in momentum.  The
Hilbert space dimension, $N=2\pi/\hbar$, plays the role of the semiclassical
parameter $kL=pL/\hbar$ in the billiard system.  The double integral of
Eq.~(\ref{vabint}) must be replaced by a double sum
\begin{equation}
S= N^2\mathop{\sum_{i,j=1}^N}_{i \ne j} \left[\overline{|\psi_i|^2|\psi_j|^2} -
c\right]^2 \,,
\label{maps}
\end{equation}
where $c$ is a constant that ensures
\begin{equation}
N^2\mathop{\sum_{i,j=1}^N}_{i \ne j} \left[\overline{|\psi_i|^2|\psi_j|^2} -
c\right] =0\,.
\end{equation}
Note that since we are working in one dimension, we must drop the $i=j$ terms
to prevent them from dominating the sum.  Our one-dimensional toy
model will not reproduce the $\ln kL/ (kL)^2$ behavior that is associated with
the short-distance $|\mathbf{r}-\mathbf{r'}| \ll L$ divergence of the two-dimensional
correlator.  Instead, we can think of $S$ as the analogue of the
two-dimensional integral (\ref{vabint}) with the short-distance part
subtracted:
\begin{equation}
V^2 \int_V \int_V d\mathbf{r} \, d\mathbf{r'} \,
C^2(\mathbf{r},\mathbf{r'}) -
{3 \over \pi} \left({2 \over \beta} \right)^2 {\ln kL \over
(kL)^2} \sim {b_g \over (kL)^2} +\cdots\,.
\end{equation}

 Numerical results for the map are shown in Fig.~\ref{figmap}.  We observe the expected $S = b_{\rm map}/N^2$ semiclassical behavior for large $N$, and the increase of the
prefactor $b_{\rm map}$ with decreasing classical stability exponent
$\lambda_0$ (see the discussion in Section~\ref{truevab}).  Furthermore, we note that even for the ``typical" case $\lambda_0=1$, strong deviations from the
simple power-law behavior appear for $N \le 50$; even larger values of $N$ are
necessary to observe the correct power law for smaller $\lambda_0$.  All the
curves saturate at $S \approx 0.045$, leading to the appearance of a slower
than $1/N^2$ decay at moderate $N$ values.  Thus, it is not surprising that a
weaker than expected dependence on $kL$ is observed for moderate $kL$ values in
Section~\ref{truevab}.

 As noted in Ref.~\onlinecite{kapalh}, the interaction
matrix element covariance is suppressed relative to the variance by a factor
$\sim kL$ or $N$, and the covariance is not a self-averaging quantity.  To
improve the poor ratio of signal to statistical noise, we may work with a
larger ensemble defined by
\begin{equation}
\label{vqrnd}
V(q)=-{q^2 \over 2} - A\cos q +V_{\rm rnd}(q)\Theta(|q|-q_0)
\end{equation}
and
\begin{equation}
\label{kprnd}
K(p)={p^2 \over 2} + A\cos p +K_{\rm rnd}(p)\Theta(|q|-p_0) \,,
\end{equation}
where $V_{\rm rnd}(q)$ and $K_{\rm rnd}(p)$ are random functions, $K_{\rm
rnd}(p)$ is even to preserve time-reversal symmetry, and $\Theta$ is the step
function: $\Theta(x)=1$ for $x \ge 0$ and $0$ otherwise.  The local dynamics
near the periodic orbit at $q=p=0$ is unaffected by the ensemble of
perturbations.  In Fig.~\ref{figcovarmap}, we use $A=0$ and $q_0=p_0=\pi/2$,
but very similar behavior is obtained for other values of the parameters.

\end{document}